\documentclass[pre,reprint,superscriptaddress,showpacs,floatfix]{revtex4-1}

\pdfoutput=1

\usepackage{natbib}
\usepackage{graphicx}
\usepackage{amssymb}
\usepackage{bm}
\usepackage{url}

\newcommand{\be}{\begin{equation}}
\newcommand{\ee}{\end{equation}}
\newcommand{\bea}{\begin{eqnarray}}
\newcommand{\eea}{\end{eqnarray}}

\renewcommand{\th}{\theta}
\newcommand{\ga}{\gamma}
\newcommand{\lam}{\lambda}
\newcommand{\ka}{\kappa}
\newcommand{\eps}{\epsilon}
\newcommand{\pa}{\partial}
\newcommand{\Om}{\Omega}

\begin{document}

\title{A domino model for geomagnetic field reversals}

\author{N.~Mori}
\affiliation{Department of Physics, Ochanomizu University, Tokyo 112-8610, Japan}
\author{D.~Schmitt}
\email[Corresponding author, email: ]{schmitt@mps.mpg.de}
\author{J.~Wicht}
\affiliation{Max-Planck-Institut f\"ur Sonnensystemforschung, 37191 Katlenburg-Lindau, Germany}
\author{A.~Ferriz-Mas }
\affiliation{Departamento de F\'isica Aplicada, Universidad de Vigo, 32004 Orense, Spain}
\affiliation{Instituto de Astrof\'isica de Andaluc\'ia, IAA-CSIC, 18080 Granada, Spain}
\author{H.~Mouri}
\affiliation{Meteorological Research Institute, Tsukuba 305-0052, Japan}
\author{A.~Nakamichi}
\affiliation{Koyama Astronomical Observatory, Kyoto Sangyo University, Kyoto 603-8555, Japan}
\author{M.~Morikawa}
\affiliation{Department of Physics, Ochanomizu University, Tokyo 112-8610, Japan}

\vskip3mm\date{\today}

\begin{abstract}
We solve the equations of motion of a one-dimensional planar
Heisenberg (or Vaks-Larkin) model consisting of a system of interacting
macro-spins aligned along a ring. Each spin has unit length and is described by
its angle with respect to the rotational axis. The orientation of the spins can
vary in time due to spin-spin interaction and random forcing. We statistically
describe the behavior of the sum of all spins for different parameters. The
term ``domino model" in the title refers to the interaction among the spins.

We compare the model results with geomagnetic field reversals and dynamo
simulations and find strikingly similar behavior. The aggregate of all spins
keeps the same direction for a long time and, once in a while, begins flipping
to change the orientation by almost 180 degrees (mimicking a geomagnetic
reversal) or to move back to the original direction (mimicking an excursion).
Most of the time the spins are aligned or anti-aligned and deviate only
slightly with respect to the rotational axis (mimicking the secular variation
of the geomagnetic pole with respect to the geographic pole). Reversals are
fast compared to the times in between and they occur at random times, both in
the model and in the case of the Earth's magnetic field.
\end{abstract}


\maketitle

\section{\label{sec:intro}Introduction}

One of the most remarkable phenomena of geomagnetism is that the Earth has
reversed the polarity of its almost dipolar magnetic field many times in the
past at irregular intervals \citep[e.g.,][]{jacobs:94,merrill:etal:96}. Similar
reversals have also been observed in turbulent dynamo experiments
\citep{berhanu:etal:07} and in simulations of the geodynamo \citep{gr:95}, but
the cause of the reversals has yet eluded a convincing explanation
\citep{amit:etal:10}.

The magnetic field of the Earth originates from dynamo action in the liquid
outer core \citep[e.g.,][]{roberts:07}. Helical flows in convection columns
that encircle the inner core tangent cylinder and are aligned with the rotation
axis play an important role for the magnetic field generation
\citep{busse:75,olson:etal:99}. Numerical simulations of the geodynamo
successfully reproduce many features of the magnetic field of the Earth
including stochastic reversals \citep[e.g.,][]{cw:07}. Depending on the
importance of the inertial forces relative to the rotational forces, dynamos
with either a dominant axial dipole or with a small-scale multipolar magnetic
field are found \citep{kc:02}. The transition from dipolar to multipolar
dynamos takes place at a local Rossby number of approximately $0.1$
\citep{ca:06}. The Earth lies close to the transition between both types
\citep{oc:06}, which may explain why the dipole undergoes sporadic reversals.

In weakly driven dynamos the helical convection columns generate an axial
dipole, while in strongly driven dynamos the flow and the field have smaller
spatial structures and chaotically fluctuate in time. Direct numerical
simulations of the geodynamo are computationally expensive and thus only a few
reversals have been studied in detail. Typically these simulations show that
reversals go along with a breaking of the north-south (equatorial) symmetry in
the flow of the aligned fluid columns. Sporadic flow upwellings, when
transporting inverse magnetic flux patterns from the inner core to the core
mantle boundaries, seem to trigger polarity reversals \citep{aubert:etal:08}.
To some extent, the upwellings behave like tilted convective columns, at least
what their role in the dynamo mechanism is concerned. We will further discuss
this issue in Sect.~\ref{sec:sim}.

The reversal sequences in paleomagnetic data and in dynamo models have been
analyzed for their statistical properties. First estimates that geomagnetic
reversals obey a Poissonian process where all reversals are independent of each
other do not describe all statistical features of the reversal record. The
statistical reversal rate has likely been changing over time due to the varying
heat flux through the core-mantle boundary
\citep{glatzmaier:etal:99,constable:00,do:09,biggin:etal:12}. The reversal
sequence also suggests that the process may have a short and a long term
memory, leading to changes in the statistical behavior and the characterising
distribution function of the times between reversals \citep{jonkers:03,rs:07}.
Similar analysis for fully 3D numerical dynamo simulations are rare because it
is very costly to compile a large number of reversals. The analysis by
\citet{wicht:etal:09} and \citet{do:09} indicate that the numerical simulations
may follow a similar reversal statistics as the paleomagnetic record.

Simple parameterized models allow for a large number of reversals so that a
statistical analysis becomes more meaningful. A famous example is the two-disk
dynamo of \citet{rikitake:58}, which exhibits sporadic reversals, but also a
cyclic variation of the dipole moment during stable polarity periods. The
extensions to $N$ coupled disks by \citet{sh:85} and \citet{ito:88} improved on
the latter weak point. \citet{hoyng:etal:01} considered a mean-field dynamo
model with stochastic fluctuations of the induction effect; these lead to
oscillations of the dipole field amplitude in a bistable potential with minima
representing normal and reversed polarity and occasional jumps between them
\citep{schmitt:etal:01}.

Here we study another class of simplified models, an Ising-Heisenberg model of
interacting magnetic spins. Ising-like models have been used in molecular
dynamics and statistical mechanics for describing, for instance, phase
transitions in ferromagnetism, for modelling spin glasses and for pattern
recognition in neural networks \citep[e.g.,][]{greiner:etal:95}. Coupled spin
models of Ising type, where the individual spins can assume two scalar states
$+1$ or $-1$ and interact with each other after certain rules, have also been
suggested for describing geomagnetic polarity reversals and their statistics
\citep{ml:89,si:93,dias:etal:08}. We analyze a planar Heisenberg model
consisting of a system of vectorial spins aligned along a ring. Each spin has
unit length and is described by its angle with respect to the rotational axis,
i.e., each spin has one degree of freedom. The orientation of the spins can
vary in time due to spin-spin interaction and random forcing. The consecutive
interaction of adjacent spins is described as ``domino model". We consider the
time dependence of the average orientation of all spins, which exhibits a
similar behavior as the geomagnetic reversal record.

This sort of models are often classified according to spatial dimensionality
and number of components of spin vectors. Our model is one-dimensional (i.e.,
spatial dimensionality of the lattice is one) and vectors (spins) are
two-dimensional (i.e., they are contained within a plane). Thus, the domino
model is a one-dimensional XY model, also referred to as the plane rotator
model or the Vaks-Larkin model \citep{vl:66}. Spin vectors in Ising and
Heisenberg models are 1D and 3D, respectively. A clear classification can be
found in the classical textbook by \citet{stanley:87}.

The spins in our model might be associated with the convection columns, whose
electromagnetic induction generates elementary dipoles. The tendency of the
spins to be aligned with the rotation axis is a consequence of the
Proudman-Taylor theorem, and the time variation of the spins is a measure of
the vigor of convection and of the sporadic upwellings. The convective columns
represent building blocks of the full dynamo process. Their interaction is
modelled here by the spins in the simple domino model. It turns out that this
model successfully describes the statistics of geomagnetic reversals, which
indicates that the polarity reversals may be understood by the collective
interaction of these columns. The model does, however, not describe the details
of the dynamo process in the individual convection columns itself.

The structure of the paper is as follows. In Sect.~\ref{sec:model} the model is
described. The results and the statistical analysis of our model are presented
in Sect.~\ref{sec:res}. The influence of the various parameters as well as
alternative model descriptions are given in Sect.~\ref{sec:par}. The results
are compared with numerical dynamo simulations in Sect.~\ref{sec:sim} and with
geomagnetic data in Sect.~\ref{sec:geo}. In Sect.~\ref{sec:dis} we give our
conclusions.

\section{\label{sec:model}The domino model}

\subsection{\label{sec:eq}Model equations}

\begin{figure}
\includegraphics[width=0.9\columnwidth]{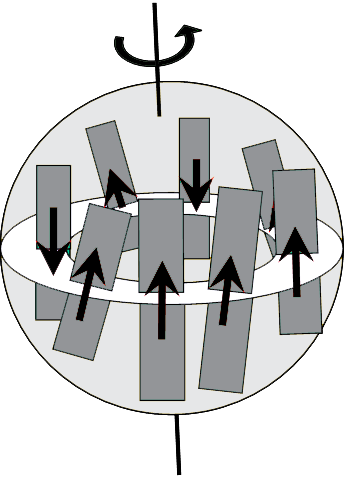}
\caption{\label{fig:model}Sketch of the domino model.}
\end{figure}

We consider a system of $N$ macro-spins aligned along a ring and interacting
pairwise like in a one-dimensional Vaks-Larkin model. The spins are embedded in
a uniformly rotating medium and we take $\bm{\Om}=(0,1)$ as the unit vector
along the rotational axis. Each spin $\bm{S}_i,\;i=1,\dots,N$ has unit length
and is described by its angle $\th_i$ with respect to the rotational axis, such
that $\bm{S}_i=(\sin\th_i, \cos\th_i$). The orientation of the spins can vary
in time due to random forcing and spin-spin interaction (Fig.~\ref{fig:model}).

The kinetic and the potential energy $K(t)$ and $P(t)$ of the system are
\bea
K(t)&=&\frac{1}{2}\sum_{i=1}^N\dot{\th}_i(t)^2 \,, \label{eq:kin} \\
P(t)&=&\ga\sum_{i=1}^N(\bm{\Om}\cdot\bm{S}_i)^2+\lam\sum_{i=1}^N
(\bm{S}_i\cdot\bm{S}_{i+1}) \,, \label{eq:pot} \eea
where $i+1=1$ when $i=N$. Here $\ga$ is a parameter characterizing the tendency
of the spins to be aligned with the rotation axis, while $\lam$ is a parameter
characterizing the spin-spin interaction. The scalar product to the square in
the $\ga$-term ensures that there is no preferred polarity. The interaction is
such that each spin interacts with the two neighboring spins: spin 2 interacts
with spins 1 and 3, spin 3 with spins 2 and 4 and so on. Spin $N$ interacts
with spins $N-1$ and 1, i.e. we are considering periodic boundary conditions
and therefore talk about a ring system here.

The Lagrangian for the system is ${\cal L}=K-P$. We set up a Langevin-type
equation as follows
\be
\frac{\pa}{\pa t}\left(\frac{\pa{\cal L}}{\pa\dot{\th}_i}\right) =
\frac{\pa{\cal L}}{\pa\th_i}-\ka\dot{\th}_i(t)+\frac{\eps\chi_i}{\sqrt{\tau}} \,,
\label{eq:lag}
\ee
where the term $-\ka\dot{\th}_i(t)$ describes friction and the term
$\eps\chi_i/\sqrt{\tau}$ is a random force acting on each spin. The parameters
$\ka$ and $\eps$ characterize the strengths of the friction and the random
forcing, respectively. Finally, $\chi_i$ is a Gaussian-distributed random
number with zero mean and unit variance associated to each spin, which is
updated each correlation time $\tau$.

Inserting the expressions for the kinetic and potential energy, (\ref{eq:kin})
and (\ref{eq:pot}), into Eq.~(\ref{eq:lag}) yields
\bea &&\ddot{\th_i}-2\ga\cos\th_i\sin\th_i+\lam[\cos\th_i(\sin\th_{i-1}
+\sin\th_{i+1}) \nonumber \\
&&\hspace*{3.5cm}{}-\sin\th_i(\cos\th_{i-1}+\cos\th_{i+1})] \nonumber \\
&&\qquad\qquad{}+\ka\dot{\th}_i-\frac{\eps\chi_i}{\sqrt{\tau}}=0\;,\;i=1,\dots,N
\label{eq:mo}
\eea
with $\th_0=\th_N$ and $\th_{N+1}=\th_1$.

We integrate the equations of motion (\ref{eq:mo}) forward in time with a
4th-order Runge-Kutta scheme starting from a random orientation of the spins
between $0$ and $2\pi$. A standard set of parameters $N$, $\ga$, $\lam$, $\ka$,
$\eps$ and $\tau$ is considered in Sect.~\ref{sec:res}, while the parameter
dependence of the mean time between reversals is studied in Sect.
\ref{sec:time}. In Sect.~\ref{sec:alt} we also slightly alter the model using
different alternatives of the $\ga$-term, the $\lam$-term and the forcing term.

As main output we consider the cumulative orientation of all spins and define
\be M(t) = \frac{1}{N}\sum_{i=1}^N\bm{\Om}\cdot\bm{S}_i(t) =
\frac{1}{N}\sum_{i=1}^N\cos\th_i(t) \label{eq:m} \ee
as the resulting total axial magnetic moment or ``magnetisation".

\subsection{\label{sec:num}Assessment of numerical stability}

We tested that the results are insensitive to the employed numerical method.
For the Runge-Kutta scheme, for instance, we reduced the integration time step
$\Delta t$ by factors of $2$, $5$ and $10$ compared to the model of
Sect.~\ref{sec:res} without qualitative and quantitative change of the results.

We furthermore implemented two different algorithms from the ODEPACK
\citep{hindmarsh:01}, a predictor-corrector scheme after Adams (suitable for
non-stiff systems) and a backward differentiation scheme after Gear (for stiff
cases). In both cases the step-size was adaptive and controlled by relative and
absolute error tolerances. Both gave the same results as the Runge-Kutta scheme
once the error tolerances were chosen small enough, e.g. relative error bounds
equal zero and absolute error bounds equal $10^{-10}$ for $\th_i$.

Although the particular times at which reversals occurred varied, the overall
statistical behavior did not change when we reduced the time step of the
Runge-Kutta routine or when we employed the other integration algorithms. Both
the mean time and the distribution of times between reversals as well as the
power spectrum and the distribution of the magnetisation were all the same
within the statistical margins.

We conclude from this that the results of the domino model are numerically
robust and we can stay with the computationally less expensive Runge-Kutta
method.

\section{\label{sec:res}Results and statistical analysis of a typical model}

\begin{figure*}
\includegraphics[width=0.9\textwidth]{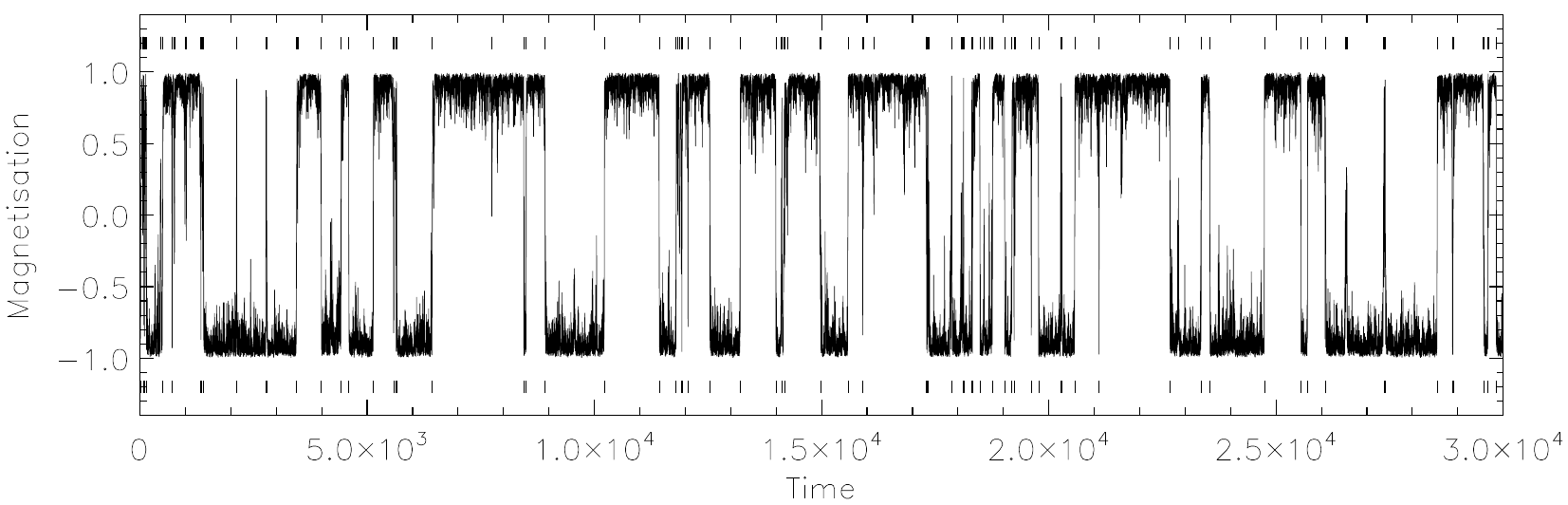}
\caption{\label{fig:mag}The magnetisation of the standard run as a function of
time. At the top the times of all zero-crossings are indicated (all reversals),
while at the bottom only those where a central band of $M=[-0.5,0.5]$ is
crossed are displayed (true reversals).}
\end{figure*}

The parameters of our standard model are $N=8$, $\ga=-1$, $\lam=-2$, $\ka=0.1$,
$\eps=0.4$ and $\tau=0.01$. The integration time step was $\Delta t=0.01$, the
total number of time steps was $3\times10^7$, and every 10th time step is
outputted. The run comprises a total of 824 reversals, i.e. the mean time
between reversals is 364. Identifying this time with the mean time between
reversals in the case of the Earth, which is 300 kyr, the whole run with a time
of $300\,000$ spans approximately 250 Myr. The length of the run is not limited
by numerical constraints, but the number of 824 reversals is large enough for a
robust statistical comparison with the available geomagnetic record of 332
reversals (see Sect. \ref{sec:geo}). In Fig.~\ref{fig:mag} the first tenth of
the full run is displayed.

\begin{figure*}
\includegraphics[width=0.9\textwidth]{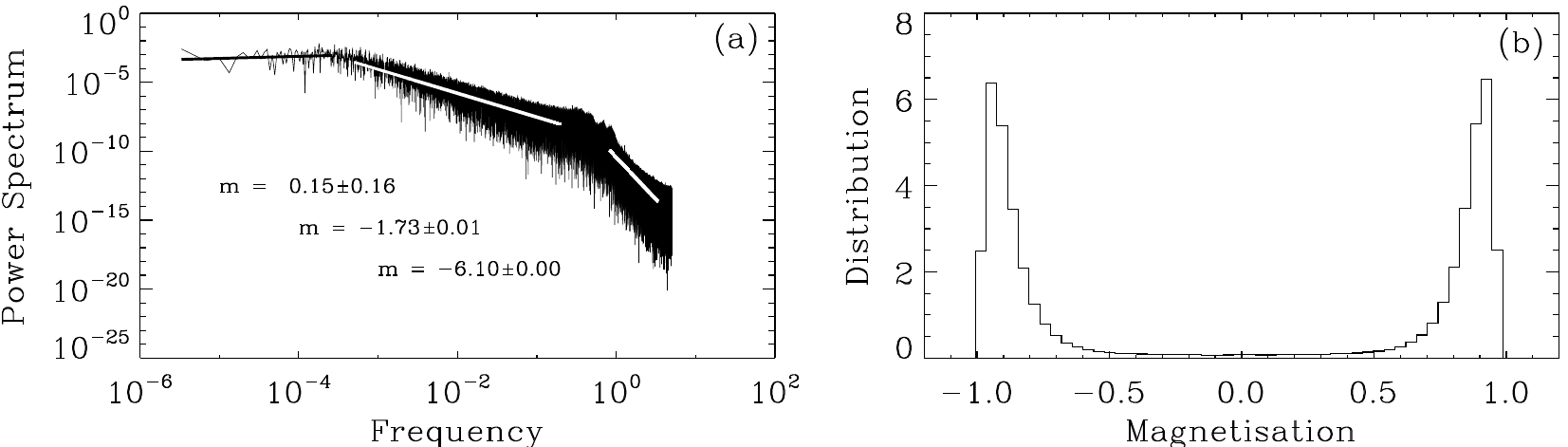}
\caption{\label{fig:power}Left (a): Log-log plot of the power spectrum of the
magnetisation. Right (b): Normalized distribution of the magnetisation.}
\end{figure*}

\begin{figure*}
\includegraphics[width=0.9\textwidth]{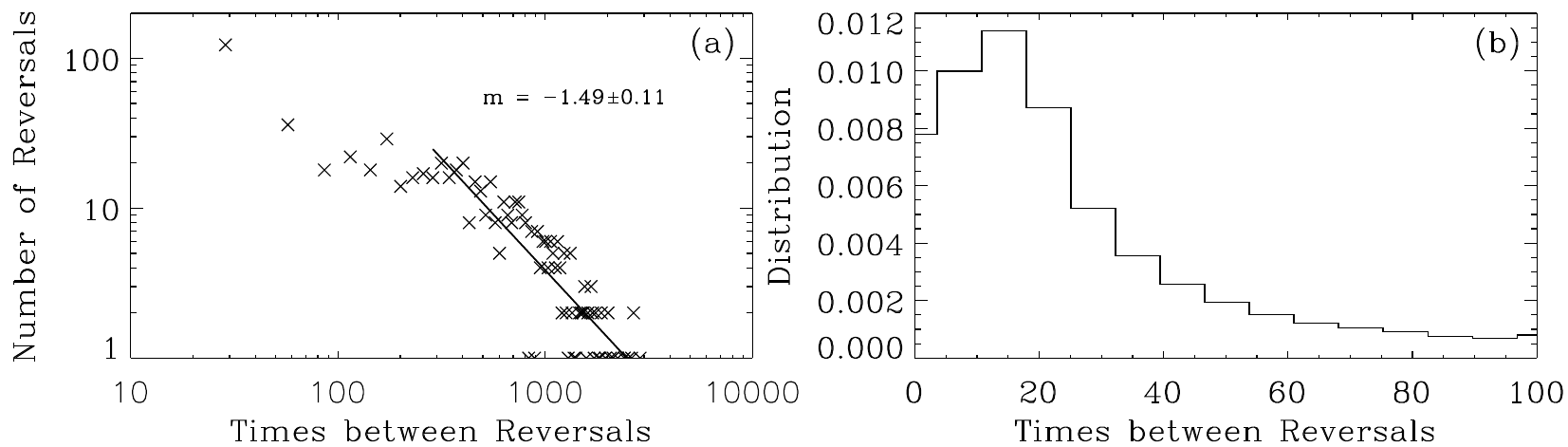}
\caption{\label{fig:dis}Left (a): Log-log plot of the number of reversals as a
function of the times between reversals. Right (b): Normalized distribution of
the times between reversals for short duration chrons.}
\end{figure*}

The statistical analysis is based on the whole time series. The power spectrum
is shown in Fig.~\ref{fig:power}a. Over a large range comprising most of the
reversals, the spectrum follows a power law with an exponent of about $-1.7$.
The spectrum for small frequencies or long polarity chrons (i.e., epochs of one
polarity) is flatter, while the steeper decrease at high frequencies comes from
the fast variations between reversals. The distribution of the magnetisation
peaks near $\pm1$ with a wide and deep valley between them
(Fig.~\ref{fig:power}b). This reflects the fact that the flipping or reversal
times are short events compared to the average duration time between them and
that the spins are most of the time closely aligned with the rotation axis.

The distribution of the duration of long chrons follows a power law with an
exponent of approximately $-1.5$ (Fig.~\ref{fig:dis}a), while the short chrons
are approximately exponentially distributed (Fig.~\ref{fig:dis}b).

\begin{figure*}
\includegraphics[width=0.9\textwidth]{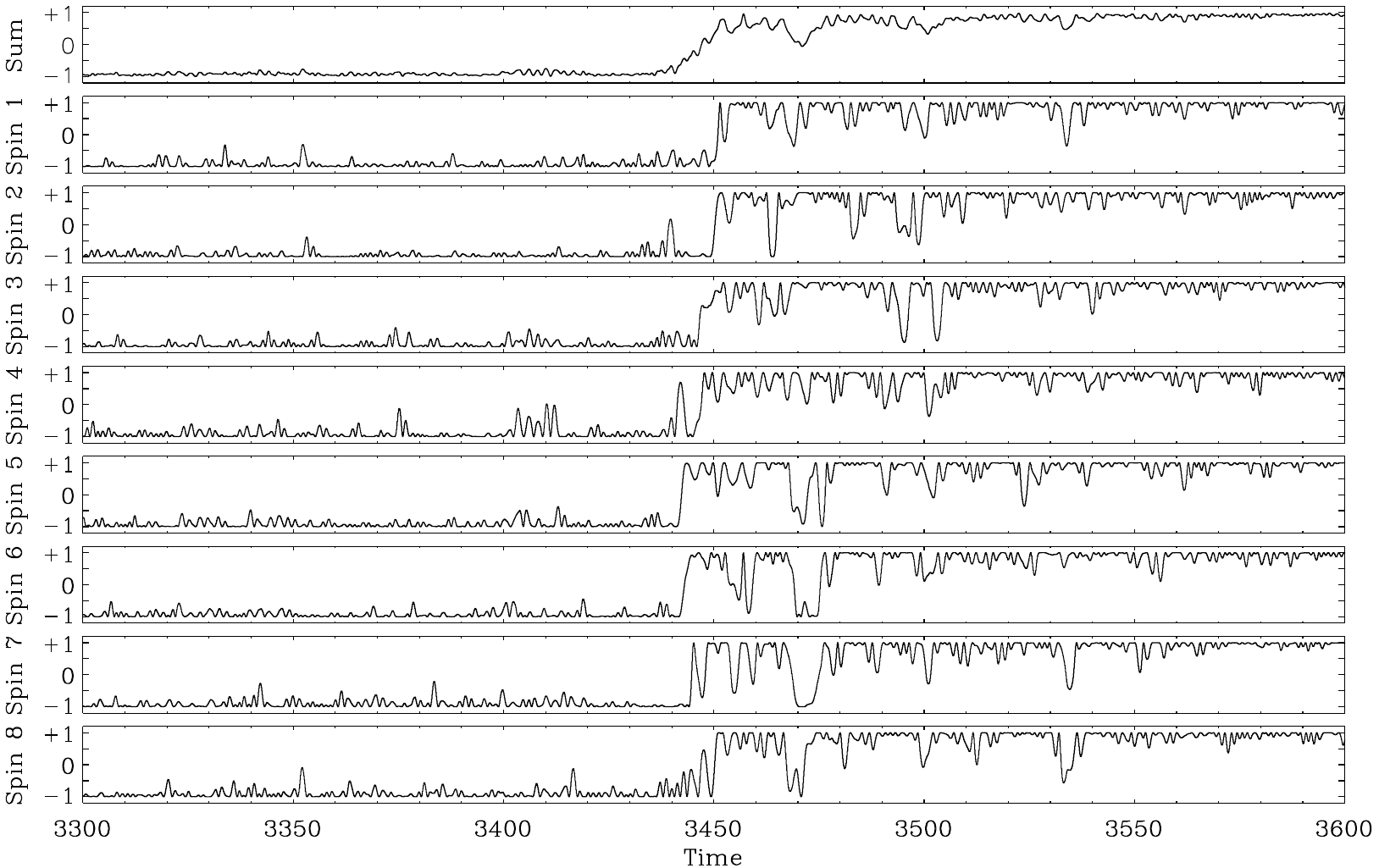} \caption{\label{fig:spins}The total
magnetisation of all spins, $\sum_{i=1}^N\cos\th_i/N$, (top panel) and the
magnetisation of the individual spins, $\cos\th_i$, before and after a reversal
at $t\approx3450$.}
\end{figure*}

\begin{figure*}
\includegraphics[width=0.9\textwidth]{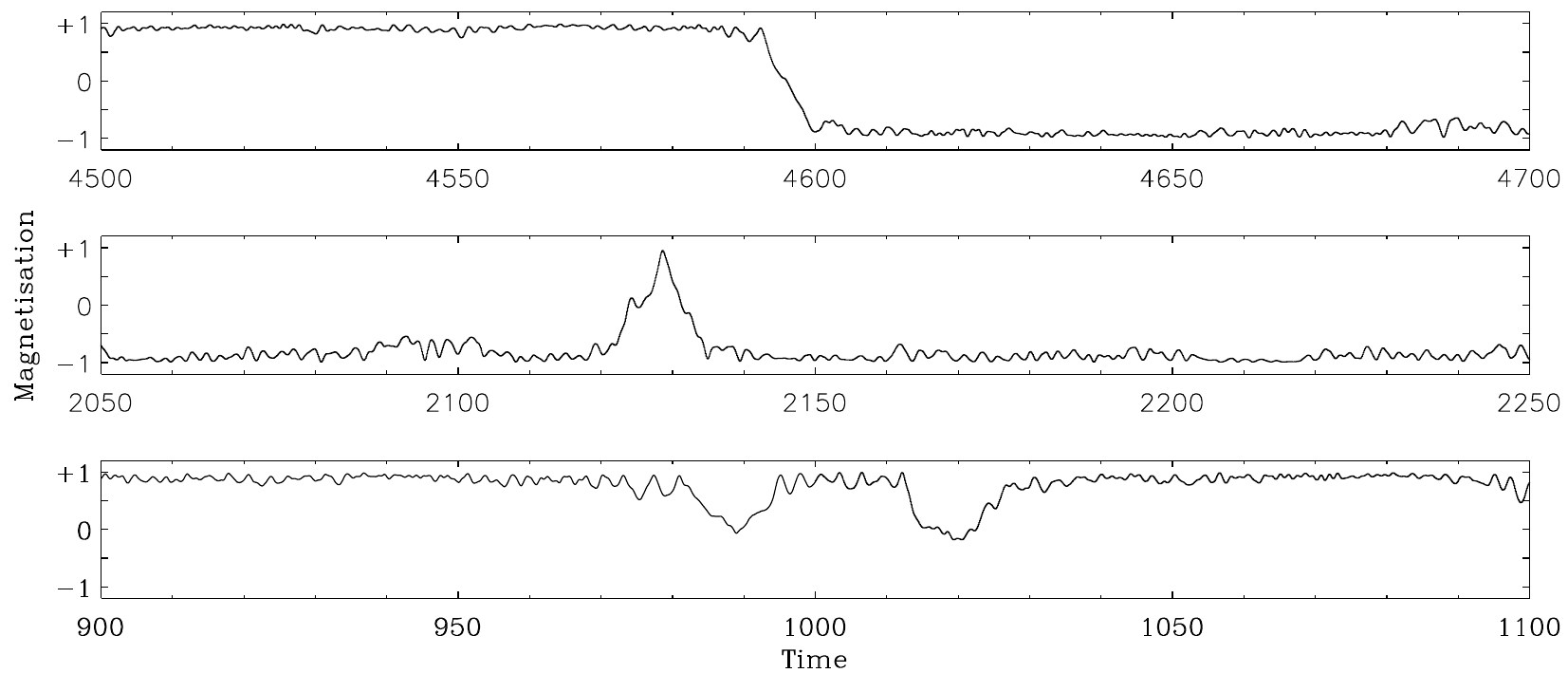} \caption{\label{fig:coll}Examples of
a true reversal (top panel), an aborted reversal (middle panel) and two
excursions (bottom panel).}
\end{figure*}

In Fig.~\ref{fig:spins} the details of one single reversal at $t\approx3450$
are shown. The reversal is triggered by a large fluctuation of spin 4 which is
successively transferred to the neighboring spins, which fluctuate until
finally all spins reverse their polarity. We call our model ``domino model"
because of this consecutive interaction of neighboring spins. The duration of
the full reversal depends on how fast the original fluctuation is transferred
to all other spins. The reversal here lasts about 10 time units, which
corresponds to 8000 yr, roughly the time it takes for the geomagnetic field to
flip polarity. Shortly afterwards, at $t\approx3470$, some spins again show
large variations and even reverse for a short time. Since they fail to transfer
this to all the other spins the total magnetisation shows an excursion rather
than a reversal.
Examples of a true reversal, an aborted reversal and two excursions are shown
in Fig.~\ref{fig:coll}.

\section{\label{sec:par}Parameter study}

The model contains a number of free parameters. The main quantity which depends
sensitively on their values is the frequency of reversals. The statistical
results, described in Sect.~\ref{sec:time}, are based on many runs with at
least three hundred reversals. When the time between reversals was longer than
in our standard model of Sect. \ref{sec:res} we also had to execute these runs
longer to achieve stable statistical results. The slope of the power spectrum
depends only weakly on the model parameters. The distribution of the
magnetisation is also similar to the model of Sect.~\ref{sec:res}. When there
are very many reversals, the valley between the two stable states is less wide
and deep.

\begin{figure}
\includegraphics[width=0.9\columnwidth]{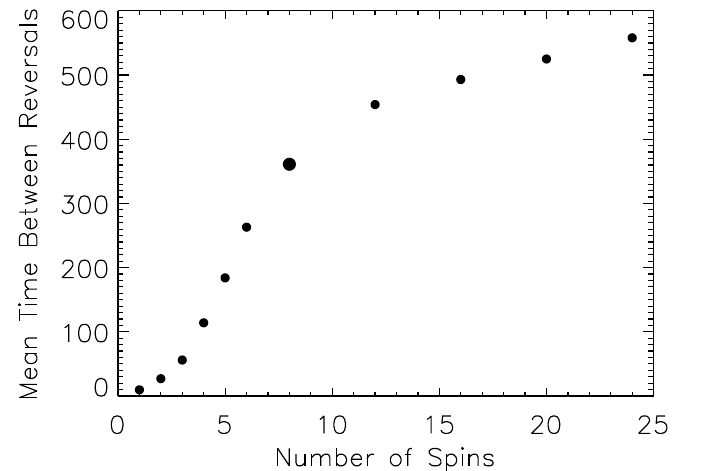}
\caption{\label{fig:n}The mean time between reversals depending on the number
of spins $N$. The symbol for the standard model of Sect.~\ref{sec:res} is
plotted larger. The values of all other parameters are as in our standard
model.}
\end{figure}

\begin{figure*}
\includegraphics[width=0.9\textwidth]{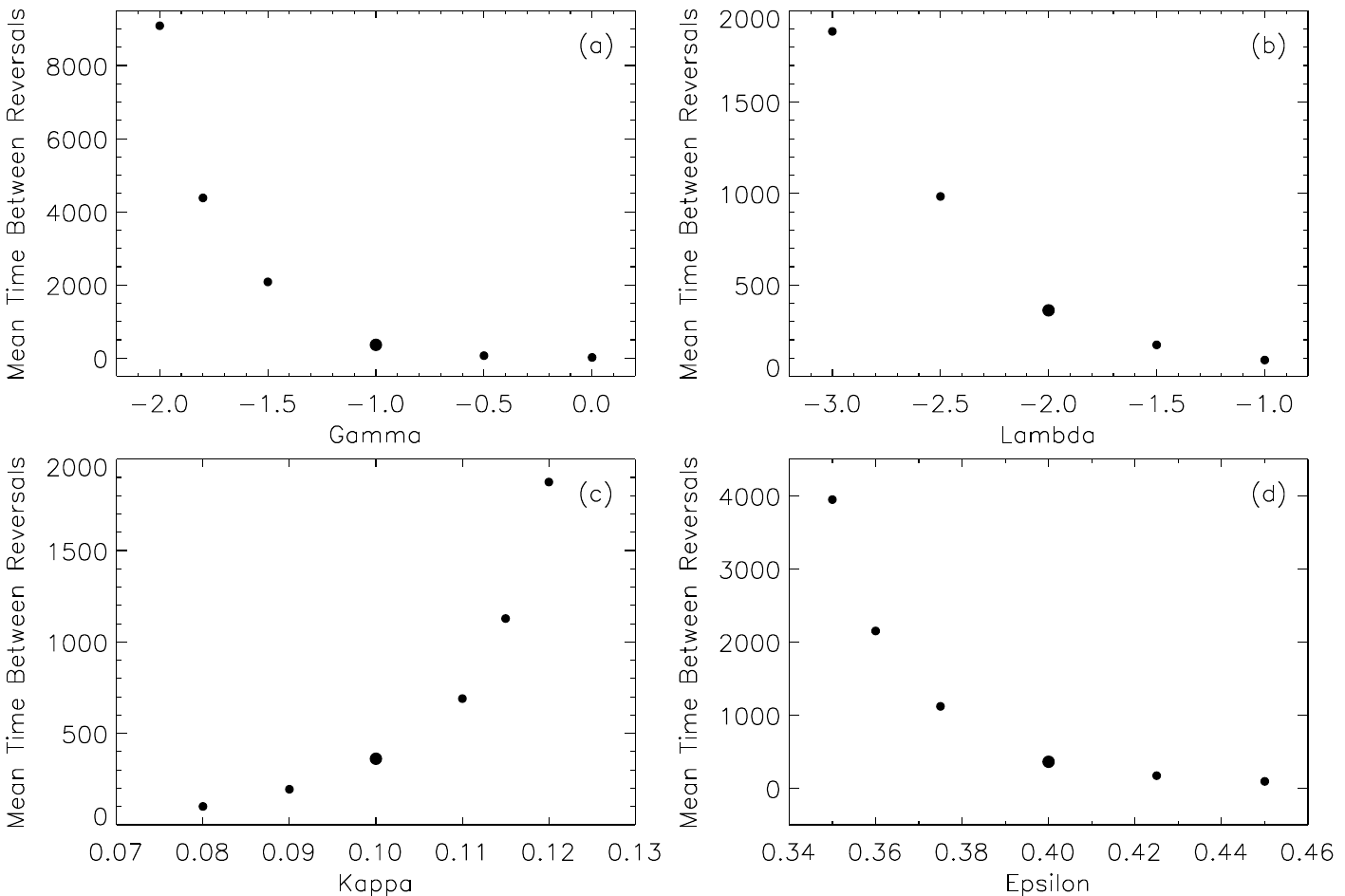}
\caption{\label{fig:par}The mean time between reversals depending on (a) $\ga$,
(b) $\lam$, (c) $\ka$, and (d) $\eps$. The symbol for the standard model of
Sect.~\ref{sec:res} is plotted larger. All other parameters are as in our
standard model.}
\end{figure*}

\subsection{\label{sec:time}Mean time between reversals}

The dependence of the mean time between reversals on the number of spins is
shown in Fig.~\ref{fig:n}. In the case of a few spins this time steeply
increases with the number, but it seems to saturate for a larger number of
spins.

The parameter $\ga$ measures the tendency of the spins to be aligned with the
rotation axis. Large negative values stabilize the orientation and lead to
fewer reversals (Fig.~\ref{fig:par}a). A value of $\ga=0$ still behaves like a
bistable oscillator, but with many reversals and an almost flooded valley
between the stable states in the distribution function. Positive values of
$\ga$ do not lead to a stable magnetisation, but to oscillations around $M=0$.

The influence of the spin-spin interaction parameter $\lam$ is similar. Large
negative values stabilize (Fig.~\ref{fig:par}b). For $\lam\ge0$ the system
randomly oscillates around $M=0$.

Increased friction, described by larger values of $\ka$, quite naturally
stabilizes (Fig.~\ref{fig:par}c), while increased random forcing, described by
larger values of $\eps$, destabilizes the system, leading to shorter chrons and
more frequent reversals (Fig.~\ref{fig:par}d).

The mean time between reversals depends sensitively on the model parameters.
Thus the drastic changes in the reversal frequency of the geomagnetic field
could be explained by a moderate change of the model parameters with time.

\subsection{\label{sec:alt}Alternative model descriptions}

We also tested to which degree the results depend on the details of the model
setup.
Instead of a $\ga$-term in Eq.~(\ref{eq:pot}) proportional to
$\sum(\bm{\Om}\cdot\bm{S}_i)^2$ we have considered a term proportional to
$\sum|\bm{\Om}\cdot\bm{S}_i|$. Since all $|\bm{\Om}\cdot\bm{S}_i|\le1$ and thus
$\sum|\bm{\Om}\cdot\bm{S}_i|\ge\sum(\bm{\Om}\cdot\bm{S}_i)^2$ very similar
results were obtained for somewhat smaller (absolute) values of $\ga$. For a
$\ga$-term proportional to $\sum(\bm{\Om}\cdot\bm{S}_i)$ no reversals were
observed.

Furthermore, instead of an additive forcing with Gaussian noise, described by
the last term in Eq.~(\ref{eq:lag}), we also investigated white noise which
only results in somewhat less frequent reversals. Using multiplicative forcing
proportional to $\cos\th_i$ leads to much fewer reversals. Multiplicative
forcing proportional to $\mathrm{mod}(\th_i,2\pi)$ yields random oscillations
about $0$ and thus not a very Earth-like reversal behavior.

As an alternative to the local interaction with neighboring spins only,
described by the $\lam$-term in Eq.~(\ref{eq:pot}), we also considered a global
or ``mean-field" interaction with all other spins, described by
$(2\lam/N)\sum_{i<j}^N(\bm{S}_i\cdot\bm{S}_j)$. A normalization factor of $2/N$
is included in order to compare with the standard interaction with just the
adjacent neighbors. The mean-field model results in less frequent reversals,
but shows otherwise qualitatively similar behavior.

It is interesting to note that in globally coupled models a qualitatively
similar behavior to the domino model described above is also found even without
noise and without friction. The resulting system is conservative and only two
parameters, $\ga$ and $\lam$, are left. We study this system in detail in
another paper \citep{nakamichi:etal:12}.

\section{\label{sec:sim}Comparison with numerical dynamo simulations}

\begin{figure*}
\includegraphics[width=0.9\textwidth]{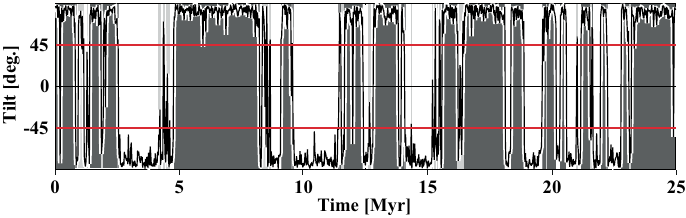}
\caption{\label{fig:sim}Dipole tilt angle in a MHD dynamo simulation with Ekman
number $E=2\cdot10^{-2}$, Rayleigh number $Ra=300$, Prandtl number $P=1$ and
magnetic Prandtl number $Pm=10$ \citep[adapted from][]{wicht:etal:09}.}
\end{figure*}

A direct comparison with numerical dynamo simulations is difficult. In our
simplified spin model the dynamic equations only concern the relative angles of
the spins to the rotation axis while a typical numerical dynamo deals with
magnetic field, velocity, pressure and temperature. Based on the close analysis
of numerical simulations
\citep{gr:95,kk:98,glatzmaier:etal:99,sj:99,kc:04,wo:04,wicht:05,
takahashi:etal:05,bouligand:etal:05,aubert:etal:08,wicht:etal:09,amit:etal:10},
see Fig. \ref{fig:sim} as an example, we may nevertheless build some analogies,
in particular concerning the reversal behavior.

Convection in fast rotating bodies like planetary liquid interiors organizes
itself in the form of convective columns. They encircle the inner core tangent
cylinder and are aligned in the $z$-direction parallel to the rotation axis.
The flow becomes quasi two-dimensional (geostrophic), minimizing any variation
in $z$-direction according to the Taylor-Proudman theorem. Cyclonic and
anti-cyclonic columns, rotating faster or slower than the planet, respectively,
alternate in azimuthal direction \citep{busse:75}. In strongly driven dynamos
the number of columns increases. In addition to this primary rotation there is
a secondary flow along the axis of the individual columns, towards the equator
in cyclonic and away from the equator in anti-cyclonic columns. Primary and
secondary component taken together determine the helicity
$\bm{U}\cdot(\bm{\nabla}\times\bm{U})$, where $\bm{U}$ is the velocity field.
For weakly driven convection (i.e., low Rayleigh number) the columns described
above dominate the flow and have helicity of one sign in the northern and of
the opposite sign in the southern hemisphere. The helicity is known to play a
crucial role for the dynamo process. The so-called $\alpha^2$-dynamo mechanism,
described by \citet{ks:97} and \citet{olson:etal:99}, can be thought of as a
process where the helicity associated to each convective column produces its
own magnetic field. The alignment with the rotation axis and the organized
helicity guarantees that the sum of these individual contributions adds up to
form the dominant axial dipole field. The dynamo equation describing magnetic
field generation can, in principle, produce field of either polarity. This
symmetry is broken, however, by the presence of a dominant dipolar background
field. Only flows with the opposite helicity in one hemisphere can weaken the
prevailing magnetic field and lead to reversals.

In order to better characterize the dynamo process in numerical simulations,
\citet{aubert:etal:08} have introduced the terms magnetic cyclone and magnetic
anticyclone. Both are directly related to the cyclonic and anti-cyclonic
convective columns. A reversal is initiated when another magnetic structure
appears that \citet{aubert:etal:08} call ``magnetic upwelling". These features
are related to radial flow upwellings and can produce inverse magnetic field.
While they seem to be a common feature at strong convective driving (i.e.,
large Rayleigh numbers) they only succeed to trigger a reversal when they are
fierce enough to also affect the field produced by the neighboring magnetic
cyclones and anti-cyclones, practically annihilating the prevailing magnetic
field. Alternatively, several upwellings can team up to do the job. Once the
background field is weak enough, convective columns are free to produce inverse
magnetic field, tilting the ``local dipole" field over and thereby convincing
neighboring columns to follow. The reversal ends once columns producing inverse
field dominate. Conceptually the magnetic upwellings can be understood as
columns with the ``wrong" helicity.

The potential energy $P(t)$ of Eq.~(\ref{eq:pot}) models two effects. Effect
one, scaled with $\ga$, is the alignment of the convective columns producing
field of either polarity with respect to the rotation axis. Effect two, scaled
by $\lam$, models the fact that the polarity that a convective structure
produces is strongly influenced by the neighboring field. The second term in
the potential energy also models the effect that a tilted spin can convince its
neighbors to follow. In the interpretation by \citet{aubert:etal:08} that would
happen once a magnetic upwelling is strong enough. Without either of these
effects, no dominant polarity or stable dipole epoch can emerge.

As the alignment of the spins with the rotation axis is enforced by increasing
the absolute value of $\ga$, we expect that the reversal rate goes down as seen
in Fig.~\ref{fig:par}a. There should be a tradeoff between $\ga$ and the random
forcing factor $\eps$. In MHD simulations this is probably related to the fact
that the Rayleigh number, and thus the convective forcing, has to be increased
to compensate a rise in rotation rate $\Om$ (i.e., decreasing Ekman number $E$)
\citep{ca:06}.

The diffusive term, scaled with $\ka$ in the Langevin Eq.~(\ref{eq:lag}),
represents the fact that a magnetic field needs time to be generated against
magnetic diffusion. Finally, the random forcing term, scaled with $\eps$ in
Eq.~(\ref{eq:lag}), encodes the flow fluctuations that seem to trigger
reversals in the simulations, for example the appearance of strong magnetic
upwellings. In kinematic dynamo models this would be described by stochastic
$\alpha$-fluctuations \citep[e.g.][]{hoyng:etal:01}.

If the friction time scale (proportional to $\ka$) becomes shorter than the
typical flow time scale (proportional to $\eps$), polarity reversals are
suppressed consistently with the behavior of the spin model (see
Fig.~\ref{fig:par}c). 
In the numerical simulations this is connected to the fact that the Rayleigh
number has to be significantly larger than the value at which dynamo action
starts. Increasing $\eps$ increases the number of reversals as shown in
Fig.~\ref{fig:par}d. From numerical simulations \citep{do:09} we know that a
stronger convective driving (i.e., larger Rayleigh number) not only leads to
more complex behavior in time and space but indeed also to more frequent
reversals.

The number of spins is an additional parameter explored here. In the MHD
simulations this is not a free parameter, but the number of columns is known to
increase with increasing $\Om$ (i.e., with decreasing $E$). Since the numerics
then becomes increasingly difficult, the effect on the reversal rate has not
really been explored so far. Statistically meaningful results only exist for
relatively small rotation rates (large $E$) \citep{wicht:etal:09}.

In conclusion, though the spin model is only a rough parametrization of the
reversal dynamics in a full 3D dynamo simulation, it nevertheless seems to
capture the main effects.

\begin{figure*}
\includegraphics[width=0.9\textwidth]{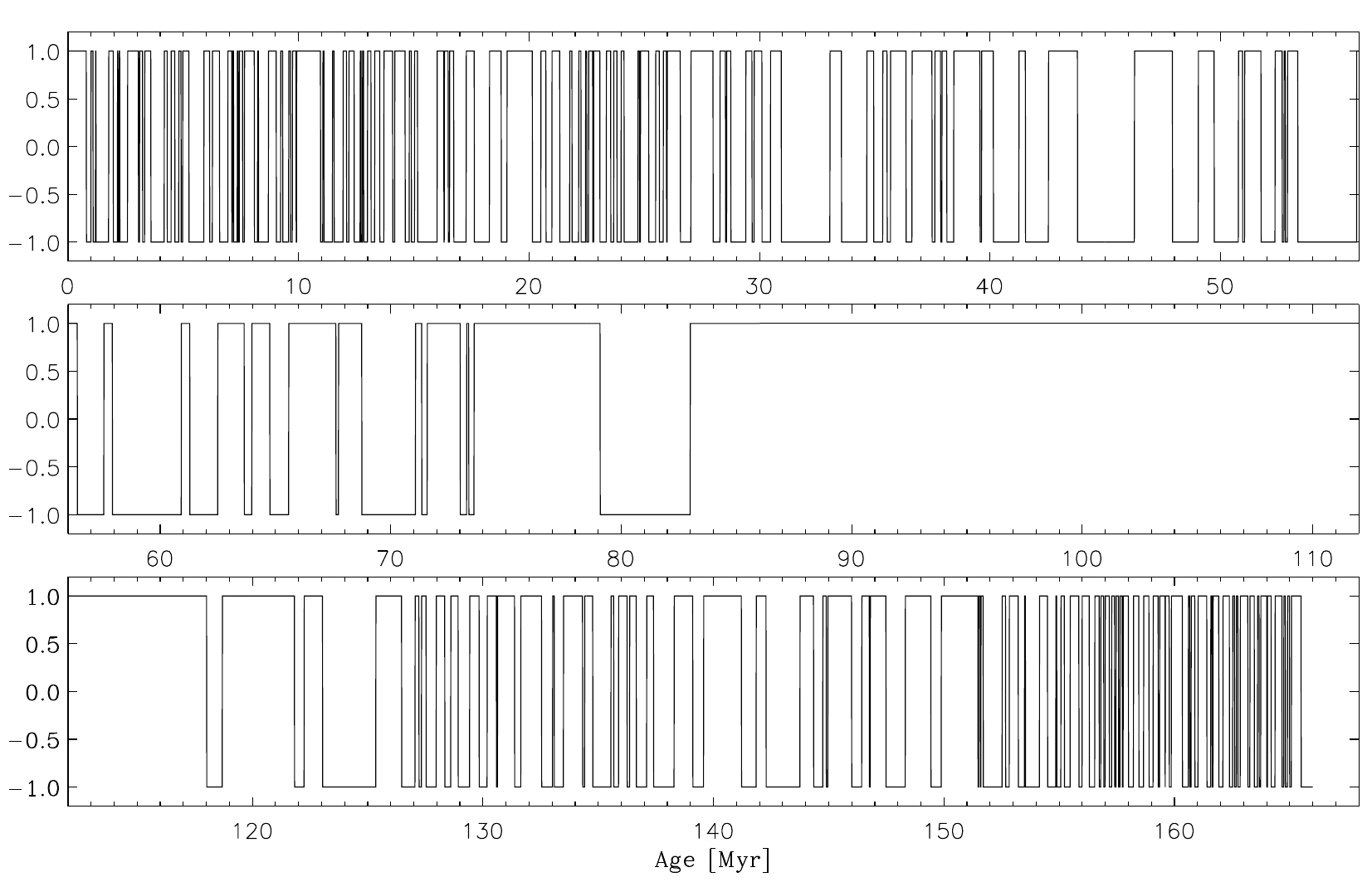}
\caption{\label{fig:geo1}Geomagnetic reversal record from present to 166 Mio yr
BP. The record comprises 332 reversals. A magnetisation of $+1$ is assigned for
chrons with normal polarity and of $-1$ for chrons with reversed polarity.}
\end{figure*}

\begin{figure*}
\includegraphics[width=0.9\textwidth]{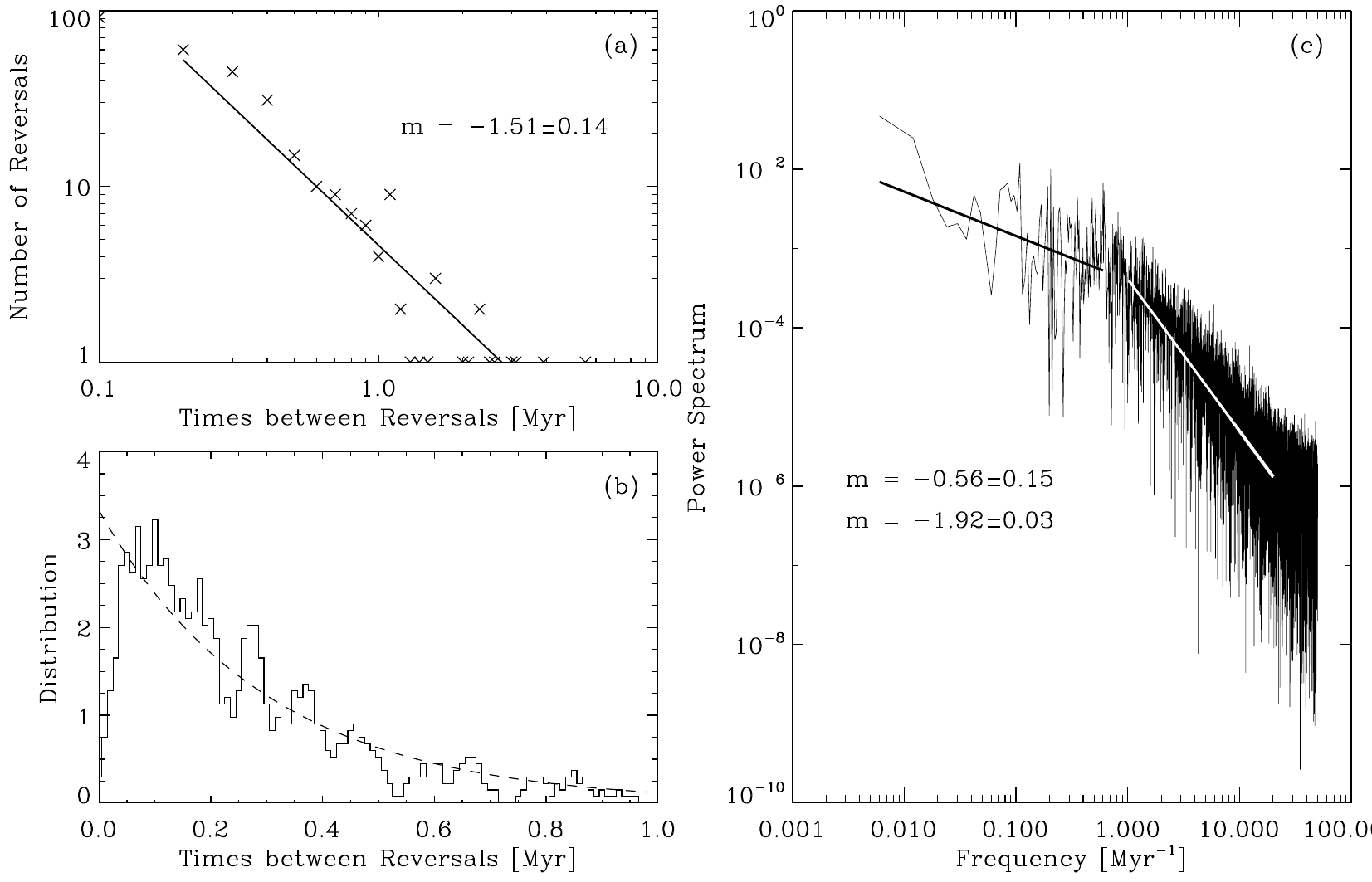}
\caption{\label{fig:geo2}Top left (a): Log-log plot of the number of reversals
as a function of times between reversals of the geomagnetic reversal record
shown in Fig.~\ref{fig:geo1}. Bottom left (b): Normalized distribution of the
times between reversals for short duration chrons of less than 1 Myr. These
make up 298 out of the total of 332 reversals. The bin size here is four times
the interval size. The dashed line is the expected probability density function
in the case of a Poissonian process with a mean polarity residence time of 300
kyr. Right (c): Power spectrum of the geomagnetic reversal record.}
\end{figure*}

\section{\label{sec:geo}Comparison with geomagnetic data}

We use the geomagnetic polarity time scale of \citet{ck:92,ck:95} and
\citet{ogg:95}, which covers the past 166 Myr and comprises 332 reversals.
Assigning a magnetisation of $+1$ for chrons of normal polarity and $-1$ for
crons of reversed polarity we derive Fig.~\ref{fig:geo1}, which is an analogue
to Fig.~\ref{fig:mag}. Reversals occurred at irregular intervals of $10^5$ to
$10^7$ yr. The mean time between reversals is approximately 300 kyr, whereas
reversals are fast events lasting only a few kyr. The reversal frequency has
considerably decreased towards and increased away from the Cretaceous
superchron which lasted from 118 to 83 Myr BP \citep{constable:00}.

The potential non-stationarity of the geomagnetic reversal record is not
present in the domino model of Sect.~\ref{sec:res}, but could be easily
accounted for by a gradual change of the model parameters with time
(Sect.~\ref{sec:par}). This may present some difficulties for the direct
comparison of the statistical analysis. The cumulative distribution of polarity
chrons roughly follows a power law with an exponent of $-1.5$
(Fig.~\ref{fig:geo2}a). Polarity intervals of a duration shorter than 1 Myr,
which make up the vast majority of all intervals, follow an exponential or
Poissonian distribution with a mean of 300 kyr (Fig.~\ref{fig:geo2}b).
\citet{rs:07} find that the full set of polarity intervals is better fitted by
lognormal and loglogistic distributions rather than Poisson and gamma
distributions \citep{constable:00}.
The power spectrum of the geomagnetic record follows power laws with an
exponent of about $-0.6$ for chrons longer than about 3 Myr and an exponent of
about $-1.9$ for chrons of shorter duration (Fig.~\ref{fig:geo2}c).

\begin{figure*}
\includegraphics[width=0.88\textwidth]{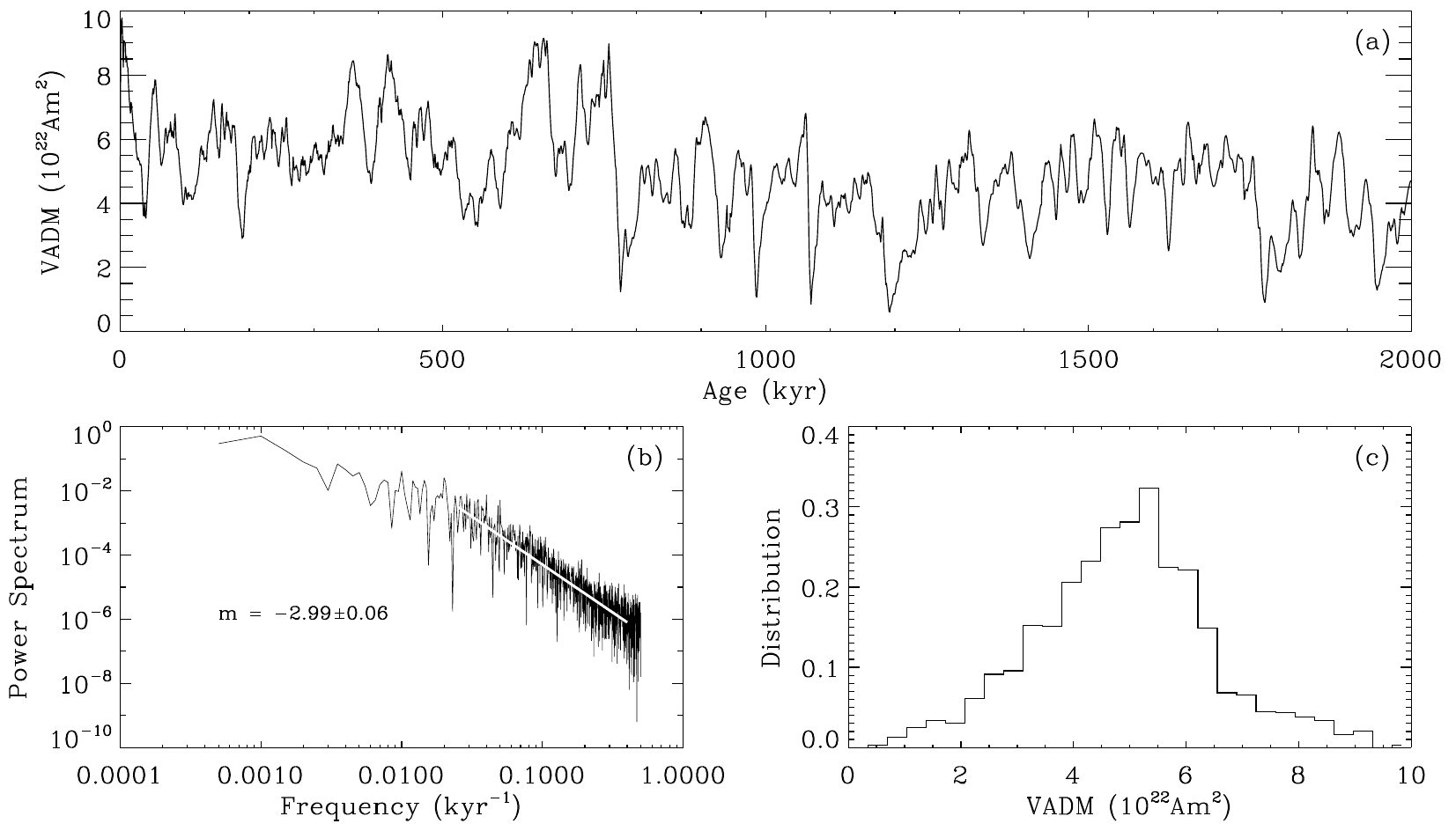}
\caption{\label{fig:sint}Top (a): Variability of the virtual axial dipole
moment (VADM) during the past 2 Myr, the SINT-2000 data set. The absolute
values of the VADM are given, disregarding the five reversals during this
period. Bottom left (b): Power spectrum of the VADM time series. Bottom right
(c): Distribution of the VADM.}
\end{figure*}

As a measure of the short-term variability of the geomagnetic field at times
between reversals we analyze the virtual axial dipole moment (VADM) of the
SINT-2000 data set (Fig.~\ref{fig:sint}a) \citep{valet:etal:05}. The power
spectrum of these fluctuations with a characteristic power index of $-3$ is
displayed in Fig.~\ref{fig:sint}b, while the distribution of the VADM is given
in Fig.~\ref{fig:sint}c. For a comparison of the distribution derived from the
shorter SINT-800 data set of \citet{gv:99}, see \citet{hoyng:etal:02}.

When comparing the geomagnetic data with the behavior of our reference model
illustrated in Figs.~\ref{fig:power} and \ref{fig:dis} striking similarities
become apparent. The fits to the chron durations suggest similar power law
exponents of approximately $-1.5$ for the paleomagnetic sequence and for our
simple model. In the latter we have disregarded the short chrons, which seem to
represent brief statistical ventures into the other polarity because they
follow a different behavior. These may be identified with paleomagnetic
excursions.

The power spectrum of the geomagnetic reversal record (Fig.~\ref{fig:geo2}c)
only refers to polarity epochs and therefore does not contain the high
frequency contributions in our model (Fig.~\ref{fig:power}c). The low-frequency
range can be interpreted as the background variation in the reversal frequency.
This leaves us with comparing the mid-frequency spectrum with a slope of $-1.9$
for the paleomagnetic data and $-1.7$ for our model. The high frequency part of
the model can be compared with the SINT data analysis, which yield a slope of
$-3$ compared to $-6$ in the model. Not surprisingly, this discrepancy suggests
that our model does a good job in replicating the statistics of the reversals,
but not in the details of the secular variation. A detailed discussion of the
power spectrum of reversals as well as intensities is presented in
\citet{cj:05}.

\section{\label{sec:dis}Discussion and conclusions}

Our simple domino model of interacting magnetic spins reproduces the
qualitative features of geomagnetic polarity reversals remarkably well. The
orientation of the aggregate of all spins is most of the time nearly aligned or
anti-aligned and deviates only slightly from the rotational axis. Once in a
while, at sporadic times, it starts flipping to ultimately change the
orientation by almost 180 degrees or to move back to the original direction.
The model thus mimics sporadic reversals of polarity, excursions and secular
variation of the geomagnetic dipole field. The power spectrum derived from the
paleomagnetic reversal records as well as the distribution of the virtual axial
dipole moment are qualitatively well represented in the model. Furthermore the
statistics of the times between reversals is similar in the model and in the
case of the Earth's magnetic field.

Our model provides a convincing statistical representation of the geomagnetic
field reversals process. One should be careful, however, when interpreting the
model properties in terms of magnetohydrodynamics. Secular variation, which is
mainly determined by the details of the convective flow dynamics, is certainly
not captured correctly. The view that the convective columns to a certain
degree represent building blocks of the full dynamo process seems to be
strengthened by our results. A stable polarity can only be established when the
majority of these entities cooperate and produce field of the same polarity.
Random forcing counteracts this and may sometimes be violent enough to cause a
spin to flip significantly and leave the team. This may cause its neighbors to
follow and ultimately lead to a global reversal. The magnetic upwellings
identified in full 3D dynamo simulations by \citet{aubert:etal:08} could be
these events. When these upwellings last long enough or produce enough inverse
field, they disrupt the normal dynamo process. The statistics of the complex
interplay of many agents seems to be nicely describable by our domino model of
Vaks-Larkin type of a set of interacting magnetic spins.

\begin{acknowledgments}
NM acknowledges financial support by the Max Planck Institute for Solar System
Research. The research of AFM has been partially funded by the Spanish {\textit
Ministerio de Ciencia e Innovaci\'on} and {\textit Ministerio de Econom\'{\i}a
y Competitividad}, through projects No. AYA2009-14105-C06-06
and
AYA2011-29833-C06,
including European FEDER funds. This work benefited from AFM's visit to the
Universities of Tokyo and Kyoto supported by a grant of the Japanese Society
for the Promotion of Science (JSPS).
\end{acknowledgments}

\end{document}